\documentclass[a4paper,10pt]{article}
\usepackage{ucs}
\usepackage[utf8]{inputenc}
\usepackage{amsmath}
\usepackage{amsfonts}
\usepackage{amssymb}
\usepackage{amsthm}
\usepackage{rotating}
\usepackage{xcolor}
\usepackage[english]{babel}
\usepackage[T1]{fontenc}
\usepackage{graphicx}
\usepackage{float}
\usepackage{hyperref}
\usepackage{cuted}
\usepackage{verbatim}
\usepackage{subfigure}
\usepackage{multirow}
\usepackage{bbold}
\usepackage{slashed}
\usepackage{indentfirst}
\usepackage{tcolorbox}
\usepackage{xcolor}
\usepackage{placeins}
\usepackage[a4paper, total={7in, 10in}]{geometry}

\usepackage{multicol}
\usepackage{fmtcount}
\usepackage{diagbox}

\usepackage{tikz}
\usepackage[compat=1.0.0]{tikz-feynman}

\newcounter{eq}

\title{\bf Charm and beauty content of
the   pion and kaon
in the Flavor U(5) Nambu-Jona-Lasinio model}
 \author{ W. F. de Sousa$^{1,2}$ , F. L. Braghin$^1$
\\
{\normalsize $^1$ Instituto de F\'\i sica, Federal University of Goias, 
Av. Esperan\c ca, s/n,
 74690-900, Goi\^ania, GO, Brazil}
\\
{\normalsize $^2$ 
Instituto Federal de Goi\'as,
R. 75,  n.46, 74055-110, Goi\^ania, GO, Brazil}
}

\date{\today}

\begin{document}

\maketitle

\begin{abstract}
We consider a U(5)-Nambu-Jona-Lasinio model with
 flavor-dependent couplings constants  obtained  from  quark-antiquark polarization
to investigate the role of the   heavy quarks  for light
constituent  quarks  and mesons.
A quantum mixing, due to the different representations
of the flavor group needed to defined quarks and mesons,
lead to charm and beauty sea quark contributions to properties of the light quark sector.
For a given  fitting procedure for the parameters of the model,
the charm and beauty quark effective masses are  freely varied 
from very small values to very large  values  (infinite).
The effect of these variations on pion and kaon observables, such 
as masses, weak decay constants and condensates,  is calculated.
Although the pion mass can vary (seemingly too much) up  to   $20\%$ 
for a very large variation of heavy quark effective masses,
their contributions for the  kaon mass  are at most  of the order of $2\%$
and 
all the other variables and observables, including up and down quark constituent masses,
vary at most around $1\%$.
\end{abstract}

\section{Introduction}

The 
quark model is a cornerstone of the Strong Interactions in the Standard Model
for which, in its standard version,
 mesons and baryons are built up respectively  with quark-antiquark and three-quark
states  \cite{quark-model,cheng-li,quark-model2}.
Besides the possibilities that arise to identify structures and effects for the valence
quarks, including diquark correlations \cite{diquark}
or  further structures
such as   tetraquarks, pentaquarks and so on \cite{tetraquarks-pentaquarks},
 quantum effects  
lead to  a
 diversity of  sea quarks and  parton structures \cite{cheng-li,weinberg-book,partons-1,partons-2,EMCg1,MS,seaquarks-qcd}.
In this respect
 sea quarks and antiquarks, sometimes encapsuled into quark condensates,
may  produce relevant contributions 
that may be probed rather in high energies
experiments
\cite{exp-N-strange,s-proton,atlas-prec,DESY-exp,s-cont-latt,nature-young}.
For example, the strange, up and down sea quark densities in the proton  
were found to be nearly the same \cite{atlas-prec}.
In spite 
of the large charm quark mass,
an intrinsic charm (IC) of the nucleon has been cogitated along the years
resulting from mixing of states
\cite{charm-1,charm-rev}.
The probability of finding a five-quark state $\left.|uudc\bar{c}\right>$ in the nucleon
was found to be around $1\%-2\%$ in different approaches, for example in
 \cite{charm-1}.
Estimation of c-content 
in  observables for other (light) hadrons have also been done \cite{light-hadrons-c,observ-c-content}.
The c-sigma term of the nucleon was calculated, for example in \cite{csigmaterm},
being that from lattice and perturbative calculations \cite{IC-lattice} 
 a matrix element $\left< N | c \bar{c} | N \right> \sim 5- 6 \%$ is obtained
whereas  beauty content would be suppressed by a factor $m_c^2/m_b^2
\sim 8 \times 10^{-2}$.
Recently, more compelling evidences were found
 for a charm content in the  proton
with 3-$\sigma$ confidence level \cite{IC-PRL,IC-nature,IC-question},
in spite of arguments against the reliability of these experimental evidences \cite{against}.
Several implications may arise for the production of  open-heavy hadrons
and other observables \cite{charm-rev}.
Different facilities and experiments have searched, started
  testing  or will investigate possible related effects, such as in Fermilab 
(D$\slashed{0}$, SELEX)   LHC, EIC, JLab and others.
 In the present work
 the emergence of contributions 
of   charm and beauty sea
 quarks are identified 
for light pseudoscalar mesons rest energies, the pion and the kaon masses.
From the experimental point of view it is not 
  currently possible to  address this question 
but it may become in the future with further theoretical investigation,
and this issue should be considered with the efforts 
 to decompose the proton - and hadrons in general - mass  from QCD
\cite{traceanomaly}.
For that, a
quantum mixing \cite{sakurai}
due to the different representations in which quarks and and quark-antiquark mesons
are defined will be considered.
This effect
  was  shown to lead to 
 a strangeness-content of the pion
 \cite{JPG-2022,PRD-2021}
and to 
 contributions for the quark-antiquark 
states of light scalar mesons  \cite{NJL-scalars}.
This mechanism will be addressed in the 
present work by considering  the   c and b quarks
in the framework of the flavor U(5) Nambu-Jona-Lasinio model.

In spite of the 
continuous efforts to produce first principles calculations  for Quantum Chromodynamics 
(QCD) it 
still is interesting to consider effective models that capture the 
relevant degrees of freedom of QCD to identify or  to 
 test particular effects or to propose new scenarios.
The Nambu-Jona-Lasinio  (NJL) model is an emblematic model for  low energy QCD  
whose structure can be directly attached to QCD and to the quark model 
\cite{NJL,klevansky,vogl-weise,hatsuda-etal,GNJL2,coimbra-etc,kondo,PRD-2014}.
It  has also been applied to heavy mesons structure with charm and beauty  quarks
\cite{NJL-heavy1,NJL-heavy2,NJL-heavy3,U5-NJL}.
One of its  most important features  is the 
description of dynamical chiral symmerty breaking (DChSB)
and its role in phenomenology with
good prediction of masses and other observables of 
several   mesons multiplets, usually and mostly  for  the light ones but not only. 
The light hadron sector is quite peculiar since approximate chiral symmetry 
and its DChSB
are considered  extremely important to define their structure and dynamics.
In this framework, the up, 
down and strange quark masses are  quite smaller than a typical 
energy scale, such as an ultraviolet cutoff or  the nucleon mass.
Other similar punctual interaction models have been considered with success
in predicting light and heavy hadron properties
\cite{SBK,mexicanos-2019,bashir-etal,adnan-etal,craig-etal}.
However, for the heavy quarks and mesons  the use of the model may be more controversial.
Being a non-renormalizable model,
an ultraviolet (UV) cutoff is needed to provide quantitative description of observables
and its values, for different regularization schemes, are quite smaller than the 
charm quark mass.
Different  regularization schemes were proposed according to which
light and heavy quarks would have either very different UV cutoffs 
\cite{SBK,mexicanos-2019,bashir-etal,adnan-etal,craig-etal}
or very similar UV cutoffs \cite{NJL-U5}.
Indeed, in this last work it has been  argued
 that  a three-dimensional cutoff 
in the non-covariant regularization 
scheme corresponds to
limit only the quark three-momentum without the quark rest energy
and in this case all quark flavors may have approximately equal  (quite low) cutoffs.
Another criticism usually raised against the NJL model for heavy quark sector, is that
the large heavy quark masses would lead to the suppression of
b and t scalar condensates 
 \cite{QCDSR-nocond}.
However, it has been shown latter that
 heavy  quark-antiquark scalar condensates - in particular the c and b -   should
not be entirely suppressed 
 \cite{NJL-t-b,anton+ribeiro}.
Nonetheless, NJL gap equations may overestimate heavy quark-antiquark scalar
 condensates \cite{NJL-U5}.

Therefore, in the present work we exploit the quantum
mixing  effect that induces  a  charm and beauty quark
content of light mesons
by means of flavor dependent coupling constants induced by vacuum polarization.
In such a U$(N_f)$ NJL-model  
the scalar-pseudoscalar interactions
 can be written in the following form:
$G_{ij}^\Gamma (\bar{\psi} \lambda_i \Gamma \psi) ( \bar{\psi} \lambda_j \Gamma \psi)$,
for $i,j = 0,1,2,..,(N^2_f-1)$ and $\Gamma=I , i \gamma_5$.
Being a strict low energy calculation, the momentum dependence
was not considered such that the non-up or down content of the pion,
and of the up and down constituent quarks, were identified mainly
for their rest energies from a bound state equation - Bethe Salpeter equation at the 
Born level - (BSE).
In \cite{NJL-U5}
 the parameters of the U(5) NJL model (current quark masses and cutoffs)
were fixed by 
fitting seven pseudoscalar mesons masses 
($\pi, K, D, D_s, B, B_s, B_c$) for a given coupling constant of reference, 
$G_0=10$GeV$^{-2}$. 
Although several SETs of parameters had been envisaged,
in the present work  one of those  SETs  of parameters  is adopted
for the numerical calculations with  degenerate up and down quark masses.
Results with the  other SETs of parameters do not change meaningfully.
As a result the scalar and pseudoscalar meson multiplets had their masses
quite well described within around $8\%$, including the $\eta's$ and the light scalar
nonet -
except the scalars $K^*_0(800)$ and $a_0(980)$.
Although  the calculation of the quark effective masses and 
of the flavor-dependent coupling constants was
 rather perturbative and  not self consistent, 
it had been shown that
the behavior of the results may be  similar in these two types of calculations 
(self consistent and non self consistent)
for the light quark sector
\cite{PRD-2021,JPG-2022}.
By freely varying the charm and/or beauty quark effective masses
their  effects on the up and down quark effective masses and on the
pion and kaon masses, and on few other observables, can be directly 
identified.
The work is organized as follows.
In the next section the model for 
 flavor-dependent coupling constants is presented
with the resulting  set of  gap equations for the quark  effective masses
or quark-antiquark condensates.
By considering degenerate up and down quarks,
without electromagnetic effects \cite{pion-kaon-em,donoghue},
the charged and neutral pion (or kaon) become degenerate.
Effects of the quantum mixing are  explicitly identified.
Some observables are discussed including
the  BSE  for the 
charged pion and kaon. 
In Section (\ref{sec:numerics})
numerical estimations are presented
by considering the parameters of the model fixed and by
varying arbitrarily the heavy quark effective masses.
In the final paragraph a summary with some final remarks.

\section{ Flavor dependent coupling constant in the  U(5) Nambu-Jona-Lasinio model}
\label{sec:model}

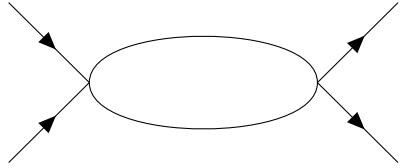
\begin{figure}[ht!]
		\begin{center}
		\begin{tikzpicture}
		\begin{feynman}
		\vertex (a);
		\vertex [right=of a] (c);
		\vertex [right=of c] (b);
		\vertex [above left=of a] (t1);
		\vertex [below left=of a] (t2);
		\vertex [above right=of b] (s1);
		\vertex [below right=of b] (s2);
		
		\diagram* {
			(t2) -- [fermion] (a),
			(t1) -- [fermion] (a)
			-- [half left, looseness=0.7, edge] (b)
			-- [half left, looseness=0.7, edge] (a),
			(b) -- [fermion] (s1),
			(b) -- [fermion] (s2),
		};
		\end{feynman}
		\end{tikzpicture}
	\end{center}
\caption{ \label{diagrams-2}
\small  
The 4-leg diagram calculated for  
  the second order terms from the large quark mass expansion of the 
quark determinant.
Numerical values are extracted for zero momentum transfer.
}
\end{figure}

The  
$U (N_f=5)$   NJL  model with flavor dependent coupling constants
generated by vacuum polarization  was derived in Ref. \cite{NJL-U5}
and it 
 is given by:
\begin{eqnarray} \label{GF-NJL}
Z[\bar{\eta},\eta] = \int {\cal D} [\overline\psi, \psi]
exp \left\{  i \; 
 \int_x \left[
\bar{\psi} \left( i \slashed{\partial} 
- m_f \right) \psi 
+ 
\frac{G_{ij}}{2}
 \left[
(\overline{\psi} \lambda_i \psi) (\overline{\psi} \lambda_j \psi) 
+ (\overline{\psi} i \gamma_5 \lambda_i \psi) (\overline{\psi} i \gamma_5 \lambda_j \psi)
\right]  
+ {\cal O} (\Delta G_{ij}^{sb})  
+ L_{s.}  \right]\right\},
\end{eqnarray}
where   the quark sources term is $L_{s.}=
\left( \overline{\eta} \psi + \eta \overline{\psi} \right)$,

$\int_x = \int d^4 x$, $i = 0, ..., (N_f^2-1)$, \ $\lambda_0=\sqrt{2/N_f} I$ and 
  $f=u,d,s,c,b$ for 
flavor in the fundamental representation.
$\eta, \bar{\eta}$ are quark sources and $m_f$ are current quark masses:
$\hat{m} = diag (
m_u, m_d, m_s, m_c, m_b )$
being considered with degenerated up and down quarks $m_u=m_d$.
The term $\Delta G_{ij}^{sb}$ stands for explicit chiral symmetry breaking 
resulting from the quark determinant that distinguishes  the scalar and pseudoscalar channels.
This method has also been applied to derive higher order 
interactions for the NJL model \cite{higher-order}.
We'll deal with the pseudoscalar meson sector 
and therefore the scalar coupling terms can be neglected, so 
that $G_{ij} = G^{ps}_{ij}$.
The corrections due to vacuum quark polarization to the NJL coupling constant
change the overall magnitude of the coupling constants:
$G_{ij}  = G_0 \delta_{ij} + \Delta G_{ij}$ where
 $G_0$ is a standard coupling constant for the standard NJL model
and  $\Delta G_{ij}$ is the correction due to polarization process 
- that is depicted in Fig. (\ref{diagrams-2}).
$G_0=10$GeV$^{-2}$
will be assumed to be a value of reference for the calculations
and 
this
induces the need of redefining the coupling constants.
Therefore the coupling constants $G_{ij}$ are the renormalized ones, usually
denoted by $G_{ij}^n$,
such that results can be directly compared to  the limit of flavor-dependent coupling constant.
Therefore it will be assumed the following
 normalization:
 \begin{eqnarray} \label{Gnorm}
G_{ij}^{n,ps} &=& G_0 \frac{ G_{ij}^{ps} }{ G_{11}^{ps} },
\;\;\;\;\;\;\;\; \mbox{such that } \;\;\; G_{11}^{n,ps} = 10 \; \mbox{GeV}^{-2}.
\end{eqnarray}
From here on, the superscript $^n$ will be omitted again.
The mixing interactions $G_{i\neq j}$ are considerably smaller than
the diagonal ones $G_{ii}$ 
\cite{NJL-U5} 
and they will be neglected in the present work
such as the effects of the  quantum mixing can be investigated  separately.

The  pseudoscalar meson fields,  $P_i$, are introduced by means of  
 the   auxiliary field method
by means of  unity integrals that multiply the generating functional given by:
\begin{eqnarray} \label{aux-variab-1}
1 &=&  N' \; \int \; D [ P_i ]
\; 
exp \;  \left[ - \frac{ i}{2 G_{ii} } \int_x \;  
  (P^i + G_{ii} (\bar{\psi}  i\gamma_5 \lambda^i \psi)_2)^2
 \right]
,
\end{eqnarray}
where  $N'$  is
a   normalization constant
and the (considerably smaller)  mixing interactions $G_{i\neq j}$
 were neglected.
From the integration of the  quark field, ground state values for the auxiliary fields
are obtained by means of saddle point equations of  the effective action.
By neglecting all the  - considerably smaller  - mixing terms these gap equations
 can be written as:
\begin{eqnarray} \label{gap-g4}
M_f - m_f  = G_{ff} \; Tr \; \int \frac{d^4 k}{(2 \pi)^4} {S}_{0,f} (k),
\end{eqnarray}
where the coupling constants $G_{ff}$  ($f=u,s,c,b$) 
were defined for the 
fundamental representation
and their relation to the $G_{ij}$ are given below.
The dressed quark propagator, with the contribution of the 
scalar quark-antiquark condensate, is:
  $S_{0,f} (k) =  \left(   \slashed{k} - M_f + i \epsilon
\right)^{-1} $.
By making $G_{ff} = 2 G_0$ one recovers the usual gap equations for the 
standard NJL model.

\FloatBarrier

\subsection{ Quantum mixing}

The  quark-antiquark interaction was initially derived in the 
adjoint representation since it is more convenient to 
define meson structure in terms of currents with the generators of the flavor group.
However the interactions are also needed to be written
in terms  of the fundamental representation
 since they are used to compute new gap equations.
For the NJL model color singlet  interactions the 
coupling constants for current-current interactions $G_{ff}$
it can be written that:
\begin{eqnarray} \label{Guu}
G_{uu} &=& \frac{2}{5}G_{0,0} + G_{3,3} + \frac{4}{\sqrt{30}}G_{0,8} + \frac{2}{\sqrt{15}}G_{0,15} + \frac{2}{5}G_{0,24} + \frac{1}{3}G_{8,8} \nonumber \\
& & + \frac{2}{3\sqrt{2}}G_{8,15} + \frac{2}{\sqrt{30}}G_{8,24} + \frac{1}{6}G_{15,15} + \frac{1}{\sqrt{15}}G_{15,24} + \frac{1}{10}G_{24,24},
\\
G_{ss} &=& \frac{2}{5}G_{0,0} - \frac{8}{\sqrt{30}}G_{0,8} + \frac{2}{\sqrt{15}}G_{0,15} + \frac{2}{5}G_{0,24} + \frac{4}{3}G_{8,8} \nonumber \\
& & - \frac{4}{3\sqrt{2}}G_{8,15} - \frac{4}{\sqrt{30}}G_{8,24} + \frac{1}{6}G_{15,15} + \frac{1}{\sqrt{15}}G_{15,24} + \frac{1}{10}G_{24,24}, \label{Gss}
\\
G_{cc} &=& \frac{2}{5}G_{0,0} - 2 \sqrt{\frac{3}{5}}G_{0,15} + \frac{2}{5}G_{0,24} + \frac{3}{2}G_{15,15} - \sqrt{\frac{3}{5}}G_{15,24} + \frac{1}{10}G_{24,24}, \label{Gcc}
\\
G_{bb} &=& \frac{2}{5}G_{0,0} - \frac{8}{5}G_{0,24} + \frac{8}{5}G_{24,24}. \label{Gbb}
\end{eqnarray}
For all these coupling constants it can be written that:
$G_{ij} = G_{ij} (M_u^*, M_d^*, M_s^*, M_c^*, M_b^*)$.
The flavor-symmetric limit is recovered immediately for equal current quark masses.
In this limit, $G_{i \neq j} = 0$, $G_{ii} = G_0$ and $G_{ff} = 2 G_0$ according to the definition above.
 From these equations
one notes  that $G_{ij}$ are calculated in terms of the corresponding quark propagators
and there are contributions of the heavy quarks polarization
into the light quark coupling constant.
The mixing coupling constants  $G_{f\neq g}$ are considerably smaller than these diagonal 
coupling constants and they will be neglected.

\subsection{ Observables and Pion and kaon masses}

As a intermediary step in the present calculation is also 
possible to provide an estimate for the heavy-quark content
of the constituent light quark (up, down and strange).
It  is seen from the gap equation (\ref{gap-g4}) that, even in the absence of mixing 
interactions $G_{i\neq j}$,  it exhibits the effect of the quantum mixing encoded in 
each of the coupling constants $G_{ff}$ that depend on all the quark effective masses.
This should be expected to correspond to
(at least a component of) a heavy quark content of the nucleon by means of 
the constituent valence quark masses.
Also all the (sea) quark flavors can be expected to contribute 
to each  chiral quark-antiquark scalar quark condensate, that 
can be written as:
\begin{eqnarray}
\left< \bar{q}q \right>_f = - \dfrac{M_f - m_f}{G_{ff}} .
\end{eqnarray}

 The Bethe Salpeter equation, or bound state equation, (BSE)
has a constant amplitude
 in the Born 
approximation \cite{NJL,klimt-etal}.
By neglecting mixing interactions, $G_{i\neq j}, G_{f_1 \neq f_2}$, in the pseudoscalar channel,
it  can be written as:
\begin{eqnarray} \label{BSE-1}
1 - 2 \; G_{ij} \; Tr \; \left(
\int_k \lambda_j  i \gamma_5   S_{0,f_1} (k + P/2) \lambda_j   i \gamma_5  S_{0,f_2}
(k - P/2) \right)  = 0,
\end{eqnarray}
where   the trace $Tr$ is for color, Dirac and flavor indices,
$Q=(M_{ps},0,0,0)$ is the meson four momentum at rest with mass $M_{ps}$.
This integral  has  quadratic and logarithmic divergences.
The indices in both adjoint and fundamental representation are needed in this equation
and they are tied by the internal structures of the bound meson states being 
respectively for charged pion and kaon $i,j=1,2$/$f_1,f_2=u,d$ and 
$i,j=4,5$/$f_1,f_2=u,s$.
Due to the non degenerate quark masses (u,s,c,b) the 
resulting equations for the  states 
can receive
two types of  unusual mixing contributions: the quantum mixing describe above
and the coupled-channel effects due to the mixing interactions $G_{f_1f_2}$.
This second type will be analyzed in another work.

In the usual chiral  limit the Lagrangian quark masses are zero,
and therefore the effective quark masses are degenerate and exclusively due to 
the DChSB.
By using the gap equations, the pseudoscalar mesons are shown to be massless 
in agreement with the Goldstone theorem.
In this case, it follows
$G_{f_1f_2} = G_{ij} = G_0$
since the flavor-dependence of the coupling constants, in the present work,
appear due to the quark mass difference.
However for the heavy quarks this is not reasonable.

The  
 weak decay
 constant 
of  charged pion and kaon
were also calculated   \cite{klevansky,vogl-weise}:
\begin{eqnarray} \label{Fps}
f_{ps} =  \frac{N_c  \; g_{psqq}}{4}\; \int \frac{ d^4 k}{(2 \pi)^4}
 Tr_{F,D} \left( \gamma_\mu \gamma_5 
\lambda_i \; S_{f_1} (k + P/2) \lambda_j S_{f_2} (k - P/2) \right)_{P=m_{ps}},
\end{eqnarray}
with the corresponding indices   $f_1,f_2$  
and $i,j$.
The meson-quark coupling constants were defined as the residue of the pole
of the BSE as:
\begin{eqnarray} \label{GMqq}
g_{psqq} \equiv  \left( \frac{ \partial \Pi_{ij} (P^2) }{\partial P_0^2 } \right)^{-2} .
\end{eqnarray}
The corresponding momentum integral calculated for on shell meson ($P^2 = M_{ps}^2$).

\section{Numerical calculations}
\label{sec:numerics}

The first part of the procedure was described with details in Ref. \cite{NJL-U5}
and the main steps will be outlined in the following.
The gap equations for the 
standard NJL-model with a reference level $G_0=10$GeV$^{-2}$
were solved
 making possible the
calculation of the flavor-dependent coupling constants 
$G_{ij}$ and $G_{f_1f_2}$.
 With these coupling constants the 
new gap equations, eqs. (\ref{gap-g4}) 
were calculated perturbatively for the corrected effective masses, $M_f$,
which will be used next to calculate meson masses and other observables.
Although we do not present a  self consistent calculation 
for effective masses  and coupling constants,
it has been shown that this perturbative
calculation  provides similar results to the fully self consistent
procedure for effective masses and coupling constants \cite{PRD-2021,JPG-2022}.
This calculation of the physical values of the  pseudoscalar meson masses
 is done firstly to fix the parameters of the model
(current quark masses and cutoffs).
After fitting the parameters, other observables can be calculated
either at the physical point (from the fitting) or with parameters/variables at different values.

In the present work we consider the SET of parameters 
fixed in \cite{NJL-U5} labeled as 
 {\bf 4} that is presented in Table (\ref{tab:parameters-gap}).
For that, 
the   current quark masses were kept  within error bars of the  values from 
Particle Data Book (PDG) \cite{PDG}.
For this SET of parameters,  the (non covariant three-dimensional) UV cutoffs
are  flavor-dependent.
Two  effective masses are shown,  $M_f$, firstly the one  obtained as solutions of the 
original gap equation (with $G_0$) and the other from the corrected gap equations (with $G_{ff}$).
Up and down quarks are taken degenerate.
The parameter $\xi_f = m_f/M_f^* $ provides the ratio of the current quark masses to the effective quark masses \cite{craig-etal}.
It measures the relative contribution of the  chiral condensate for the corresponding 
effective quark mass that is $1 - \xi_f$.
 The change in this parameter  when using $G_{ff}$ instead of $G_0$ is larger
for heavy quarks.

The  corresponding values of observables  and quantities  
at the physical point  for the light hadrons are written in Table (\ref{table:PSMesMas-M1}),
the remaining of the observables 
for heavy mesons were provided in \cite{NJL-U5}.

\begin{table}[ht]
	\caption{ \small 
		Set of parameters, $m_f, \Lambda_f$, and 
solutions of the gap equations for
standard NJL with $G_0$ and for flavor-dependent NJL coupling constants
$G_{ff}$ from Ref. \cite{NJL-U5} - SET {\bf 4}.}
	\centering \label{tab:parameters-gap}
	\begin{tabular}{| c | c | c c c c | }  \hline \hline  
		Set 			& $f$ 				& $u$ 	& $s$ 	& $c$  	& $b$  \\ \hline \hline 
		4 			& $m_f$ (MeV)       & 9.0 	& 150 	& 1250	& 4600
		\\
						& $\Lambda_f$ (MeV)	 
& 530   	& 530 	  & 540  	& 540   
		\\
		($G_0$)			& $M_f$ (MeV)       & 470 	& 667 	& 1873 	& 5236 
		\\
		($G_{ff}^{ps}$)	& $M^*_f$ (MeV)     & 466 	& 634 	& 1786 	& 5121
		\\
						& $\xi_f$ 	($G_0$)		& 0.019 & 0.225 & 0.667 & 0.879
		\\
    					& $\xi_f$ ($G_{ff}^{ps}$)	& 0.019  & 0.237  & 0.699   & 0.898
		\\ \hline \hline 
	\end{tabular}
\end{table}
\FloatBarrier

\begin{table}[ht]
		\caption{
 Light meson observables for the sets of parameters 
			 shown in Table (\ref{tab:parameters-gap}):
outside parenthesis calculation done  with
$G_0$ and inside parenthesis
$G_{ij}$ and  $G_{ff}$.
The experimental or expected value (lattice QCD) are also shown (Exp.).
$^\dagger$The  pion mass displayed is an averaged value of the neutral and charged masses.
$\bar{S}_f = (-\left< \bar{q}_fq_f \right>)^{1/3}$.
			} 
		\centering 
		\begin{tabular}{l   c c  c c c c c} 
			\hline\hline 
$m_{\pi}$ (MeV)   & $m_{K}$ (MeV)   &  $g_{\pi qq}$	&  $g_{Kqq}$	& $f_{\pi}$ (MeV)	 
 & $f_{K}$ (MeV)	  &  $\bar{S}_u$ (MeV) & $\bar{S}_s$ (MeV)  
 \\
			\hline \hline
			\\ 
 165 (140)  &    515 (503)  &  3.79 (3.61) 	&  4.03 (3.80) 	& 122 (121)
 &   123 (124)    &  359 (358) &     373 (371)
\\
  137$^\dagger$  &  495   & -  & - &    92  	&   110 	 
	   &   283 \cite{aoki-decay,lightcondensate} 	&  
290 \cite{aoki-decay,lightcondensate}
			\\
			\hline
\hline
		\end{tabular}
		\label{table:PSMesMas-M1} 
	\end{table}
\FloatBarrier

\subsection{ C and B content of pions and kaons}
\label{sec:cbcontent}

In this section
some   observables/quantities of the pion and of the kaon 
 will be calculated by varying freely 
the charm and beauty quark effective  masses.
They will be varied separately, $M_c$ or $M_b$,
  and also simultaneously.
Results for these three cases  will be presented in the same figures so that
in the horizontal axis of the figures below  there is a ratio
$\bar{M}_h/M_h$ (h=c,b)
 of a 
variable heavy quark mass,
$\bar{M}_h$,
  with respect to the corresponding  {\it physical } 
heavy quark mass
$M_h$ found in the procedure of fixing  parameters.
The third curve, with the variation of both heavy quark masses simultaneously,
was obtained by calculating the coupling constant with each of the 
corresponding values of $M_c$ and  $M_b$ from the other two curves.
 Correspondingly the calculated quantities or observables exhibited in the next figures,
 such as $ G_{uu}, G_{ss}, M_\pi, M_K$, without a 
bar represent the values at the physical point.
The same  variables or observables with a bar, such as
$\bar{G}_{uu}, \bar{G}_{ss}, \bar{M}_\pi, \bar{M}_K$, 
represent 
to  resulting quantities when varying $M_h^*$ for $h=c, b$ 
freely.

\begin{figure}[H]
    \centering
    \begin{minipage}{0.49\textwidth}
        \centering
        \includegraphics[width=1.0\textwidth]{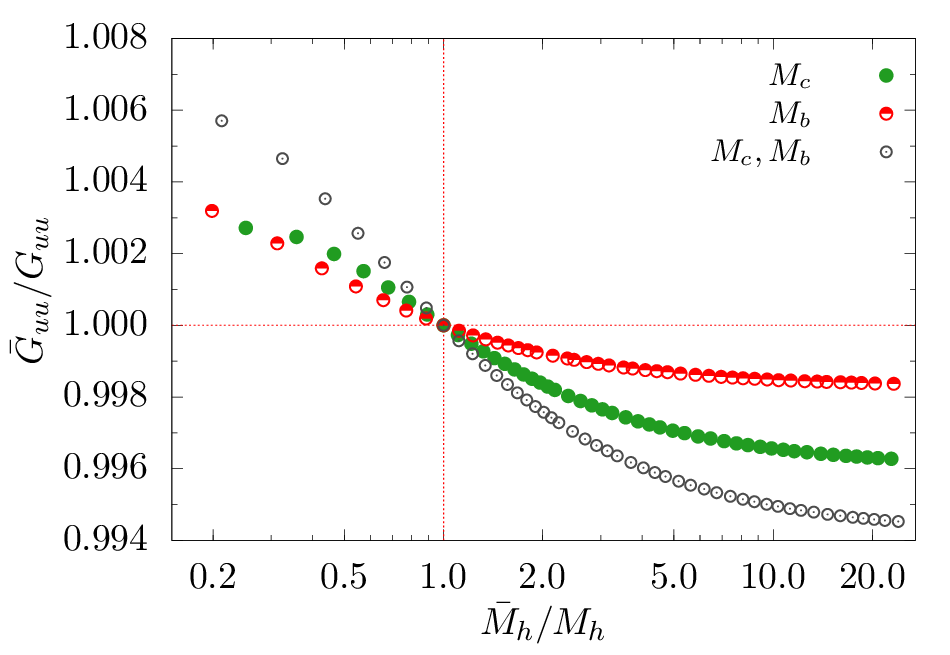}
\caption{
The  up quark-antiquark normalized coupling constant $\bar{G}_{uu}/G_{uu}$, eq. (\ref{Guu}),
  as a function
 of  ${\bar{M}_h^*}/M_h^*$  ($h=c,b$ and both) arbitrarily varied.
}
        \label{fig:Guu}
    \end{minipage}\hfill
    \begin{minipage}{0.49\textwidth}
        \centering
        \includegraphics[width=1.0\textwidth]{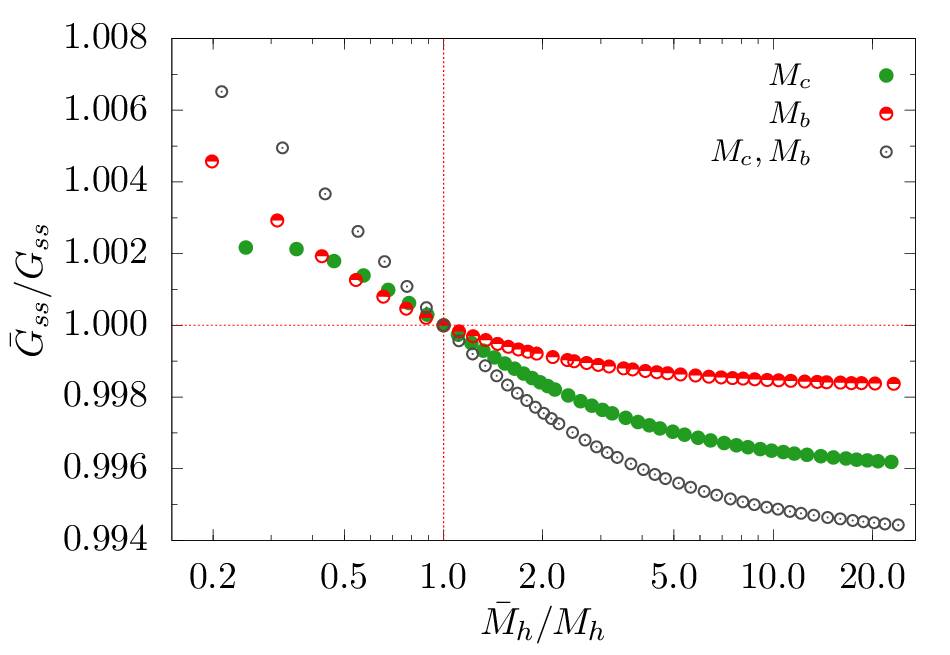}
\caption{
The  strange quark-antiquark
 normalized coupling constant $\bar{G}_{ss}/G_{ss}$, eq. (\ref{Gss}),
  as a function
 of  ${\bar{M}_h^*}/M_h^*$ ($h=c,b$ and both) arbitrarily varied.
}
        \label{fig:Gss}
    \end{minipage}
\end{figure}

In Figs. (\ref{fig:Guu})
and   (\ref{fig:Gss}) the up and strange 
quark-antiquark normalized coupling constants
are presented by means of the ratios $\bar{G}_{uu}/G_{uu}$ and $\bar{G}_{ss}/G_{ss}$
  as a functions
 of ${\bar{M}_h^*}/M_h^*$, $h=c, b$ and both
effective  masses simultaneously  varied.
The values of $\bar{G}_{ff}$ decrease with the increase of c and b effective masses.
It is seen that the maximum variation of the coupling constant with the 
large variation of values of 
c and b sea quarks  
is less than $1\%$.
We note rather that the limits
 of infinite $M_h\to \infty$ 
and zero $M_h \to  0$ exhibit opposite indications of
the role of the heavy quark masses leading to respectively weaker and 
stronger up (or down) and strange coupling constants.
This is similar to the strangeness content in the up-down quark sector of the
NJL- model, although a smaller effect because  c- and b- masses  
are much larger.
In spite of the decreasing $\bar{G}_{ff}$ with increasing heavy quark masses $\bar{M}_h$,
for very low (totally unphysical) b and c effective masses 
 there is a change in this behavior. 
The coupling constants in this range of $\bar{M}_h$
decrease for heavy quark effective masses of the order of magnitude
of nearly $\bar{M}_h \sim M_f$ (for $f=u,s$).
This is seen   for the case of varying $\bar{M}_c$ in the figures
but it is not shown for varying $\bar{M}_b$ because
the bottom quark effective mass is much larger than 
$M_f$ and the corresponding limit
is not exhibited.
This behavior will manifest in all the observables calculated with these $G_{ff}$
as seen  below.

\FloatBarrier

\begin{figure}[H]
    \centering
    \begin{minipage}{0.49\textwidth}
        \centering
        \includegraphics[width=1.0\textwidth]{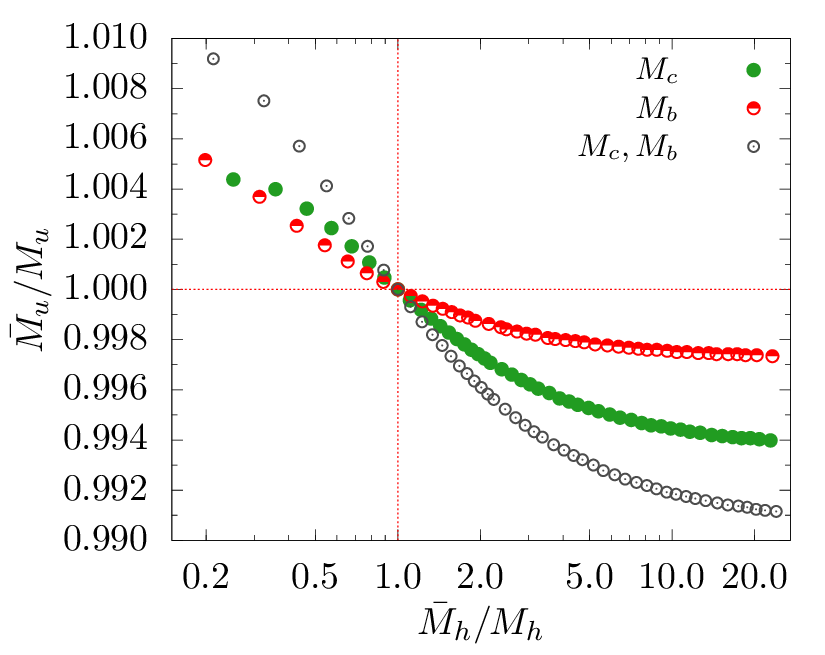}
\caption{
The behavior 
of the 
 up quark effective mass, represented in the dimensionless ratio  $\bar{M}^*_u/M_u^*$, 
calculated in  eq. (\ref{gap-g4}),
  as a function
 of ${\bar{M}_h^*}/M_h^*$ ($h=c,b$ and both) arbitrarily varied
with the same notation of the previous figures.
}
        \label{fig:Mustar}
    \end{minipage}\hfill
    \begin{minipage}{0.49\textwidth}
        \centering
        \includegraphics[width=1.0\textwidth]{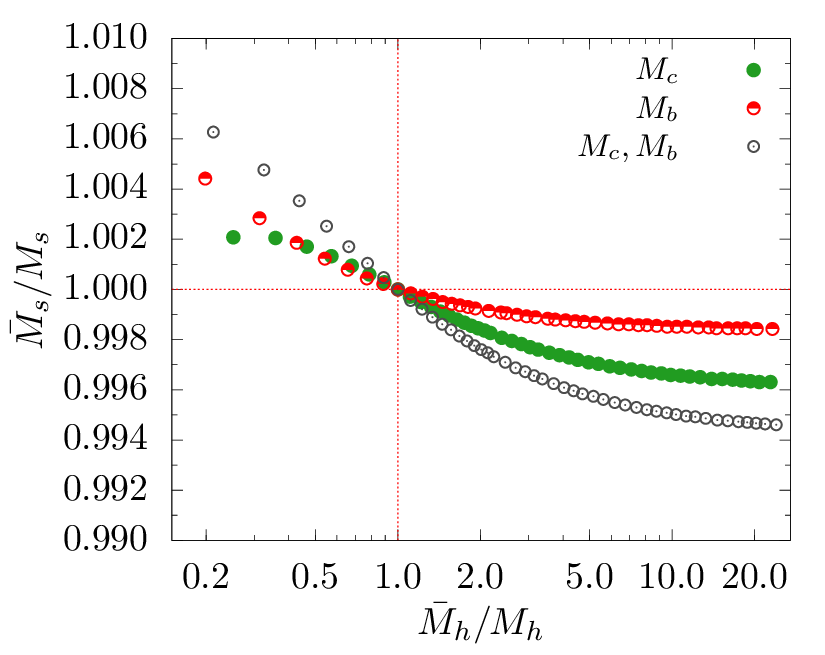}
\caption{
The  ratio of the strange quark effective mass  $\bar{M}^*_s/M_s^*$, 
calculated in  eq. (\ref{gap-g4}),
  as a function
 of ${\bar{M}_h^*}/M_h^*$ ($h=c,b$ and both) arbitrarily varied
with the same notation  of the previous figures.
}
        \label{fig:Msstar}
    \end{minipage}
\end{figure}

In Figs. (\ref{fig:Mustar})
and (\ref{fig:Msstar}) the  behavior of the  up and strange
quark effective masses  $\bar{M}_u/M_u$ and $\bar{M}_s/M_s$
is  plotted as functions 
 of  ${\bar{M}_h}/M_h$,  for $h=c,b$ and both,  arbitrarily varied.
These curves were obtained from the gap Eqs. 
  (\ref{gap-g4})
by considering the coupling constants  presented respectively  in Figs. (\ref{fig:Guu})
and  (\ref{fig:Gss}).
Due to the shape  of the gap equations the behavior of $M_u$ and $M_s$
for increasing c and b effective masses is the same as those
presented in the previous figures for the coupling constants $\bar{G}_{uu}$ 
and $\bar{G}_{ss}$.
Relative contributions reach  $1\%$
 for very large and simultaneous variations of $\bar{M}_c$ and 
$\bar{M}_b$.
 There is a tendency of the up and strange quark masses to decrease (oscillate)
for very low (totally unphysical) b and c effective masses ($\bar{M}_h  \sim M_f$ for 
$f=u,s$)
that have the same origin as discussed above for the coupling constants $G_{ff}$.

\FloatBarrier

\begin{figure}[H]
    \centering
    \begin{minipage}{0.49\textwidth}
        \centering
        \includegraphics[width=1\textwidth]{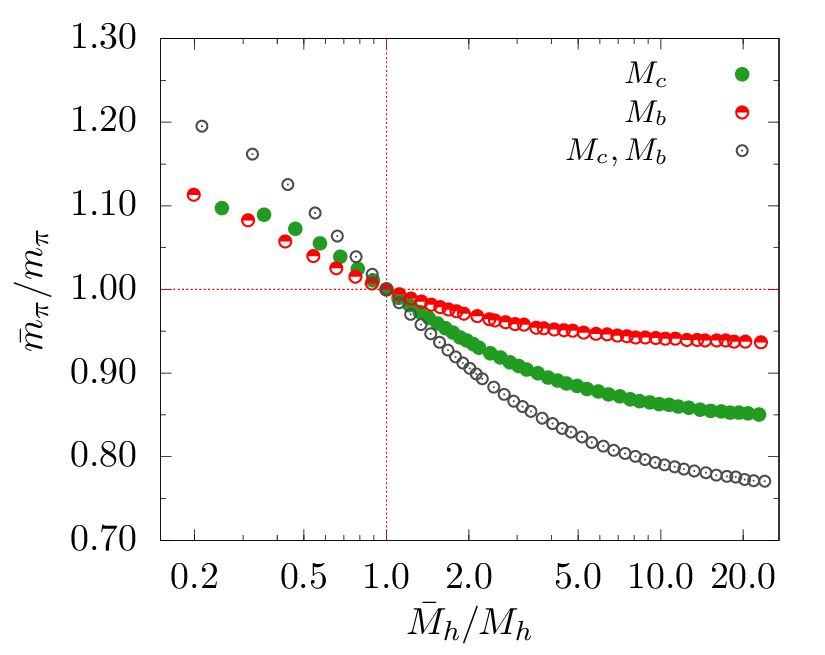}
\caption{
The  ratio of the  charged pion mass, $\bar{m}_\pi/m_\pi$,
calculated in  eq. (\ref{BSE-1}),
  as a function
 of ${\bar{M}_h^*}/M_h^*$ ($h=c,b$ and both) arbitrarily varied
in the same way of the previous figures.
}
        \label{fig:Mpi}
    \end{minipage}\hfill
    \begin{minipage}{0.49\textwidth}
        \centering
        \includegraphics[width=1\textwidth]{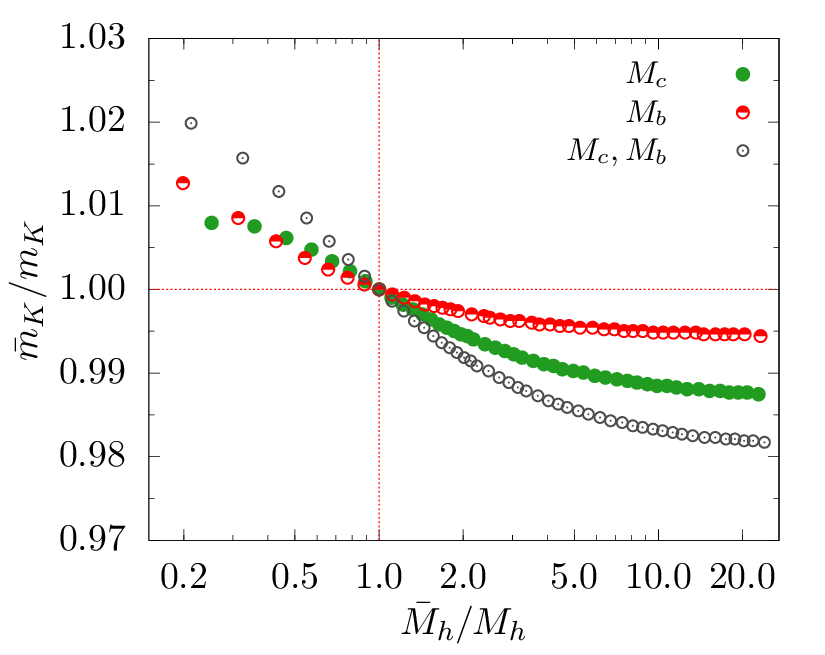}
\caption{
The charged kaon mass  $\bar{m}_K/m_K$
calculated in  eq. (\ref{BSE-1}),
  as a function
 of  ${\bar{M}_h^*}/M_h^*$  ($h=c,b$ and both) arbitrarily varied
in the same way of the previous figures.
}
        \label{fig:Mkaon}
    \end{minipage}
\end{figure}

In Figures (\ref{fig:Mpi}) and  (\ref{fig:Mkaon})
the  pion and kaon masses, represented in the
ratios of the varying charged pion and kaon masses to their 
{\it physical} values, $\bar{M}_\pi/M_\pi$
and   $\bar{M}_K/M_K$,
calculated with Eq. (\ref{BSE-1}),
are  plotted as functions 
 of  ${\bar{M}_h}/M_h$,  for $h=c,b$ and both,  arbitrarily varied.
For that, the coupling constants presented in Figs. (\ref{fig:Guu})
and (\ref{fig:Gss}) together with the 
varying quark effective masses
 were used.
Note however that  the corresponding $G_{11}, G_{22}$
 and $G_{66},G_{77}$,
respectively for the pion and for the kaon,  are used in the BSE,  being that 
they are combinations of the $G_{ff}$ shown above.
  As a consequence of the structure of the  BSE 
the masses also decrease as $\bar{M}_h$ increases.
Because of the behavior of up and strange quark effective mass 
for  very low charm effective mass, there is also 
a slight decrease of $\bar{M}_\pi$ and $\bar{M}_K$ at very low $M_c$.
However, whereas the pion mass varies up to around $20\%$ for the whole
heavy quark effective masses, 
note that the kaon mass varies only up to around $2\%$.
The most important reason for the
larger  variation in the pion mass (than in the kaon mass)
 is
the relative contribution of the heavy quark masses $M_h$ 
for the up (and down) quark effective mass.
 It can be directly noted that
 $$\frac{\Delta_{c,b} M_u}{M_u }> \frac{\Delta_{c,b} M_s}{M_s},$$
and a similar relation for the up and strange coupling constants $G_{11},G_{22}$
and $G_{44},G_{55}$ in the BSE for the pion and kaon respectively.
This has several effects  in each of the pion and kaon BSE 
(via $M_f$, $G_{ff}$ and in the momentum  integration)
such that the pion BSE receives larger corrections.
For the sake of comparison 
the variation of the pion mass with the strange quark effective mass
was also found to reach around $25\%$ in Ref. \cite{JPG-2022}.

\FloatBarrier

\begin{figure}[H]
    \centering
    \begin{minipage}{0.49\textwidth}
        \centering
        \includegraphics[width=1\textwidth]{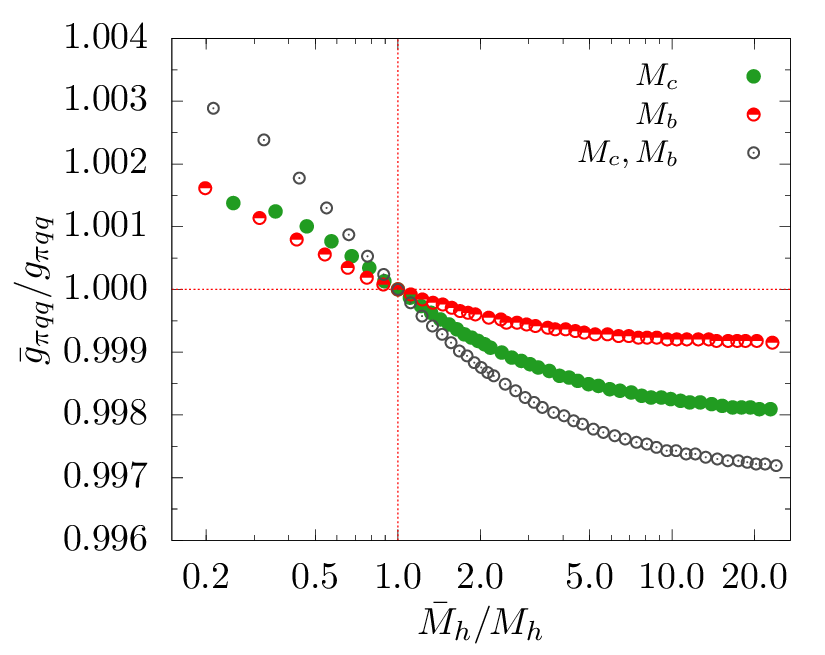}
\caption{
The the  pion-quark coupling constant, $g_{\pi qq}$, 
calculated with Eq. (\ref{GMqq}),
  as a function
 of ${\bar{M}_h^*}/M_h^*$ ($h=c,b$ and both) arbitrarily varied
in the same way of the previous figures.
}
        \label{fig:gpiqq}
    \end{minipage}\hfill
    \begin{minipage}{0.49\textwidth}
        \centering
        \includegraphics[width=1\textwidth]{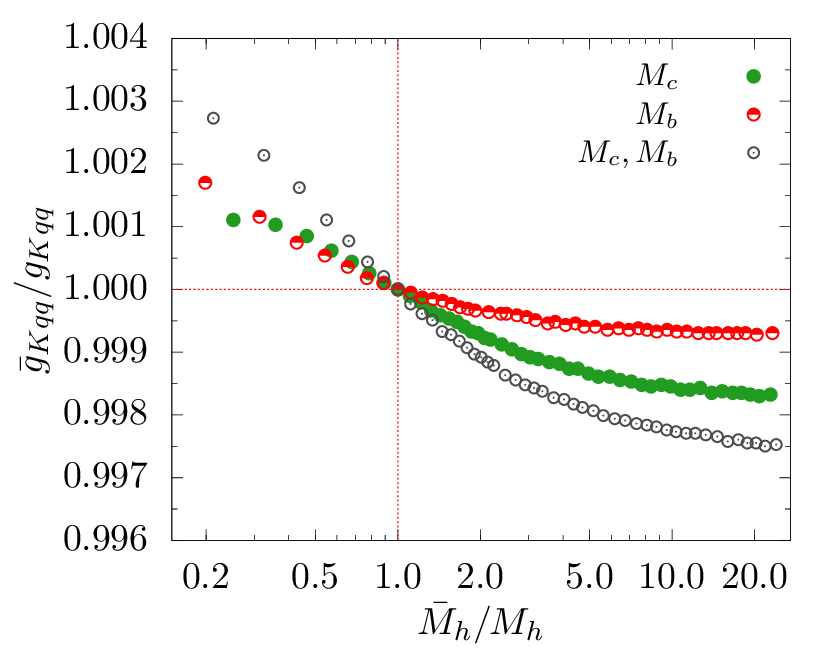}
\caption{
The the  kaon-quark coupling constant, $g_{K qq}$, 
calculated with Eq. (\ref{GMqq}),
  as a function
 of ${\bar{M}_h^*}/M_h^*$ ($h=c,b$ and both) arbitrarily varied
in the same way of the previous figures.
}
        \label{fig:gkaqq}
    \end{minipage}
\end{figure}

In Figures  (\ref{fig:gpiqq}) and  (\ref{fig:gkaqq})
the  pion-quark coupling constant
and the kaon-quark coupiling constant, $g_{\pi qq}$
and  $g_{K qq}$,  as
calculated with Eq. (\ref{GMqq}),
 are  plotted as  functions 
 of  ${\bar{M}_h}/M_h$,  for $h=c,b$ and both arbitrarily varied.
The decrease of these coupling constants
manifest the behavior of the previous variables when
varying $\bar{M}_h$.
For the largest variations of the heavy quarks effective masses
$\bar{M}_h$ the resulting coupling constants vary
at most around $0.3\%$, that is very small.

\FloatBarrier

\begin{figure}[H]
    \centering
    \begin{minipage}{0.49\textwidth}
        \centering
        \includegraphics[width=1\textwidth]{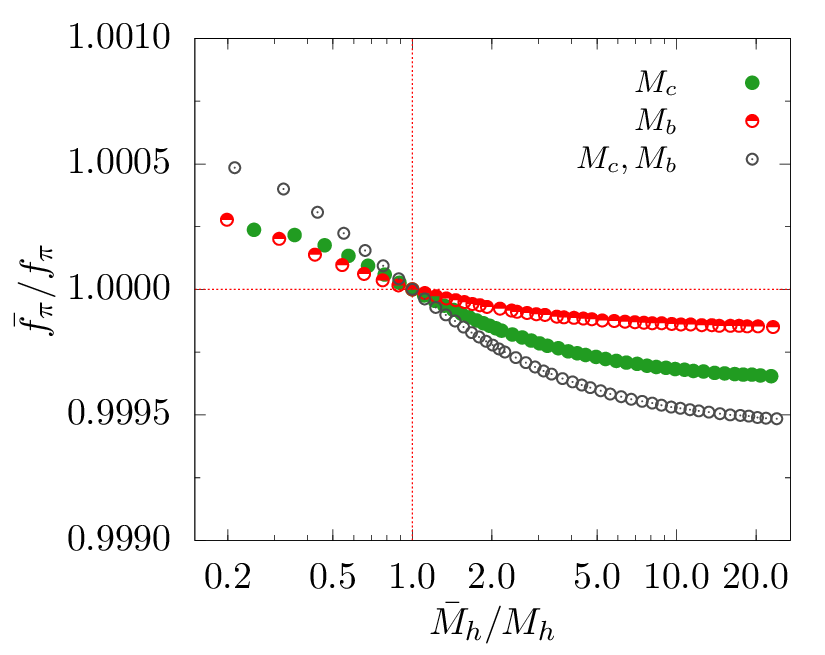}
\caption{
The  pion decay constant constant, $f_{\pi}$, 
calculated with Eq. (\ref{Fps}),
  as a function
 of ${\bar{M}_h^*}/M_h^*$ ($h=c,b$ and both) arbitrarily varied
in the same way of the previous figures.
}
        \label{fig:fpi}
    \end{minipage}\hfill
    \begin{minipage}{0.49\textwidth}
        \centering
        \includegraphics[width=1\textwidth]{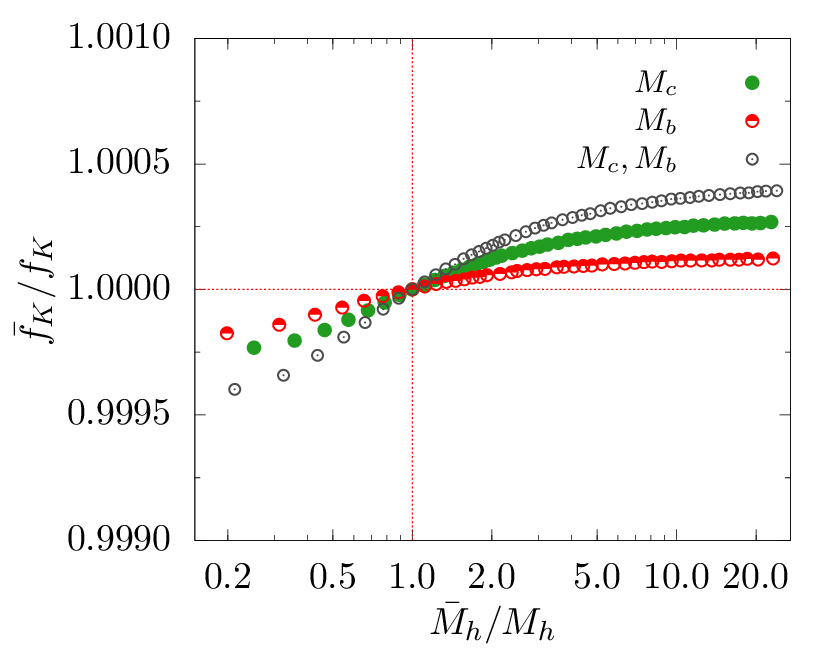}
\caption{
The  kaon decay constant constant, $f_{K}$, 
calculated with Eq. (\ref{Fps}),
  as a function
 of ${\bar{M}_h^*}/M_h^*$ ($h=c,b$ and both) arbitrarily varied
in the same way of the previous figures.
}
        \label{fig:fK}
    \end{minipage}
\end{figure}

In Figures (\ref{fig:fpi}) and  (\ref{fig:fK})
the  ratio of the  charged pion weak decay constant
and of the charged kaon weak decay constant
with respect to their values in the physical point, $\bar{f}_\pi/f_\pi$
and  $\bar{f}_K/f_K$, are
presented according to Eq. (\ref{Fps})
 as a functions 
 of  ${\bar{M}_h}/M_h$,  for $h=c,b$ and both arbitrarily varied.
 These  decay constants present an extremely  small variation.
Although  $f_\pi$ 
 decreases with increasing values of the heavy quark effective masses,
the kaon weak decay constant however presents the 
opposite behavior.  
The reason is the larger pion mass variation 
than the kaon mass variation with the increase of the heavy quark  effective masses
as noted above
$\frac{\Delta_{c,b} M_u}{M_u }> \frac{\Delta_{c,b} M_s}{M_s}$.
Being the kaon mass  considerably larger than the pion mass,
$M_\pi^2/M_K^2 \sim 1/10$,
in the equation for $F_{ps}$ Eq. (\ref{Fps})
the kaon mass has nearly the inverse behavior of the strange quark effective mass
and of the meson-quark coupling constant $g_{psqq}$.
The relative  variation of the decay constants is smaller than  $0.1\%$ for the 
large range of variations of the heavy quark effective masses.

\FloatBarrier

Several  values of observables calculated for specific values of the 
heavy quark effective masses are exhibited 
in Table (\ref{table:observ}).
Results from columns 1, 2 and 5 were taken from Tables of Ref.  \cite{NJL-U5}
and they help 
to assess the   c- and b- content of the mesons (or rather their behavior) 
together with the other columns.
Results from column 1 were obtained in the Physical point for the fitting of the SET
of parameters while column 2 presents results just by exchanging $G_{ij}\to G_0$ 
for the SET of parameters of column 1.
Therefore these two columns show the relative contribution of the 
coupling constants $G_{ij}$ which makes possible the analysis of the 
c- and b- content of the light hadrons observables.
It is clear that the flavor dependent coupling constants lower the 
meson masses.
Column 5 presents results by switching off the heavy quark condensates 
with effective masses reducing to the corresponding current masses.
This limit provides, in general, an intermediary value for the observables
when comparing with results from column 1 ($G_{ij}$) and column 2  ($G_0$).
Column 3 presents results for  very low (unphysical)  c and b quark effective masses
$M_c = 0.2 M_c$ and $M_b = 0.2 M_b$ that are considerably lower than their 
current masses.
Column 4 exhibits the limit of very large (infinite)
 heavy quark effective masses that seemingly is a (reasonable) way of
freezing the heavy quark degrees of freedom in the light hadrons.
Consistently with  the figures above, the observables reach a constant value 
for heavy quark effective masses below
 $\bar{M}_h/M_h  \sim 20$.
The overall variation of each of the observables, as the heavy quark effective masses vary 
over a large range of values, is very small except for the pion mass and 
 for the kaon mass, as discussed above.

\begin{table}[ht]
		\caption{
 Values of some observables taken at different specific values of 
$M_h$ discussed above.
Results from columns 1, 2 and 5 were taken from Tables of Ref.  \cite{NJL-U5}.
			} 
		\centering 
		\begin{tabular}{c|c c c c c} 
			\hline\hline 
Observ.  			& 1- Phys. ($G_{ij}$) 		& 2-  Phys ($G_0$) 		&  3- 
$0.2 \times (M_c,M_b)$ &  4- $(M_c,M_b) \to \infty$ & 5-  $M_c=m_c, M_b=m_b$
\\
\hline
$m_{\pi}$ (MeV)   	& 137.80	& 167.40	& 164.70	& 104.40	& 144.40		
\\
$m_{K}$ (MeV)   	& 503.20	& 515.00	& 513.20	& 493.60	& 505.20		
\\
$f_{\pi}$ (MeV)	 	& 121.66	& 121.72	& 121.71	& 121.59	& 121.67		
\\
 $f_{K}$ (MeV)	  	& 123.10	& 123.33	& 123.05	& 123.15	& 123.08		
\\
 $M_u$ (MeV) 		& 465.66	& 470.40	& 469.94	& 461.34	& 466.64		
\\
$M_s$ (MeV)  		& 634.32	& 666.72	& 638.30	& 630.74	& 635.08		
			\\
			\hline
\hline
		\end{tabular}
		\label{table:observ} 
	\end{table}
\FloatBarrier

\section*{ Final remarks}

In this work we exploited few consequences of the 
quantum mixing that emerges
 when identifying that
quarks and mesons belong to  different representations of the 
flavor group by means of the NJL model with 
flavor dependent coupling constants.
The quark effective masses and flavor dependent coupling constants
 were calculated perturbatively instead of self consistently,
being that  it has been shown for the light quark sector (flavor SU(3))
results are similar \cite{JPG-2022}.
 These coupling constants, both diagonal and mixing interactions,
can be said to
  lead to mixings between all the quarks
$U_{f_1,f_2}$ for $f_1,f_2=u,d,s,c,b$, besides
the lowering of quark effective masses. 
Mixing effects
 may have consequences that usually are analyzed with the  
CKM matrix.
This NJL-model based  calculation might overestimate the 
quark condensates although gap equations of the NJL model
may receive further contributions that have not been considered.
The effect of the  mixing interactions
$G_{i\neq j}$ and $G_{f_1 \neq f_2}$, 
that are considerably smaller than 
the diagonal $G_{ii}$, in the gap and BSE 
were not considered
being  outside the scope of the work.
The overall heavy quark dependencies of some light hadron observables,
 $g_{\pi qq}, g_{Kqq}$ and $f_{\pi}, f_K$,
 and related quantities,
$M_u, M_s, G_{uu}, G_{ss}$,
are in general   smaller than $1\%$ for a very large variation of 
the heavy quark effective masses - $0.2 M_h^{phys} < \bar{M}_h < \infty$ where
$h=c,b$ and $M^{phys}_h$ is the c- or b- mass at the physical point.
This very low variation is of the order of magnitude of
 earlier estimations for different observables
calculated with different methods
 \cite{seaquarks-qcd,charm-1,light-hadrons-c}.
The exceptions to these small variations
  are  the  variation of the pion mass,
that reached around $20\%$, and of  kaon mass up to around $2\%$.
This variation of the pion mass is nearly of the same order of magnitude
of the variation of the pion mass due to the variation with strange quark effective mass
analyzed in
\cite{JPG-2022}.
These are too large variations that signal that
the calculation for this model most probably overestimates
the c- and b- content of the rest energies from the
BSE that was considered - at the Born level.However,
 in general, the variation of the observables
with $M_c$ and $M_b$ is one or two orders of magnitude lower than 
their variations with $M_s$ found in \cite{JPG-2022}
because $M_s^2/M_c^2 \sim 1/10$ and $M_s^2/M_b^2 \sim 1/100$.
There are different ways of assessing the non-valence quark (in the present case
the c and b quarks) contributions to the
pion/kaon masses with different physical meanings
as it can be noted from Table (\ref{table:observ}).
For example,
one could consider a heavy quark contribution for the pion mass as the 
difference of the pion/kaon mass calculated in two limits:
the regular physical  pion/kaon  mass  (in Table (\ref{table:observ})) and its value when
 the heavy quark masses go close to zero zero.
In that limit  there is a slight increase of most of the light meson 
(light quark) observables
- except $F_K$.
In the opposite regime, for very large/infinite heavy quark effective masses, 
when heavy quark degrees of freedom should be frozen,
there is a slight decrease of 
most of the light meson (light quark) observables
- except $F_K$.
The calculations of the 
c-sigma term and b-sigma term contain some ambiguities  when considering
$G_{ij}$ and they were not done in the present work.
Pseudoscalar and scalar mesons mixings induced by 
the flavor-dependent interactions will be 
analyzed in another work.


\section*{Acknowledgements}

F.L.B. is member of
INCT-FNA,  Proc. 464898/2014-5,
and  he acknowledges partial support from 
 CNPq-312750/2021-8.
The authors thank short discussions with C.D.Roberts,  T. Frederico.

\end{document}